\documentstyle[12pt]{article}
\def\be{\begin{equation}}
\def\ee{\end{equation}}
\def\lb{\label}

\def\R{\hbox{\bf R}}
\begin{document}

\title{Interrelations between Quantum Groups and
Reflection Equation (Braided) Algebras.}

\author{A.P. Isaev\thanks{e-mail address: isaevap@theor.jinrc.dubna.su}\\
Bogolubov Laboratory of Theoretical Physics, JINR,  \\
Dubna, 141 980, Moscow, Russia }

\date{}

\maketitle

\begin{abstract}
We show that the differential complex $\Omega_{B}$ over
the braided matrix algebra $BM_{q}(N)$
represents a covariant comodule with
respect to the coaction of the Hopf
algebra $\Omega_{A}$ which is a differential
extension of $GL_{q}(N)$. On the other hand, the algebra
$\Omega_{A}$ is a covariant braided comodule
with respect to the coaction of the braided Hopf algebra
$\Omega_{B}$. Geometrical aspects of these
results are discussed.
\end{abstract}

\newpage

{\bf 1.} In the present letter we demonstrate the intimate
interrelations between two famous $q$-algebras: quantum groups
(matrix pseudo-groups) \cite{FRT,Wor} with the defining relations
\be
\lb{1}
\R\, T\, T' = T\, T'\, \R
\ee
and reflection equation (or braided) algebras (see \cite{KS}-\cite{Ma3}
and references therein) with
the commutation relations
\be
\lb{2}
\R \, u\, \R\, u = u\, \R \, u \, \R \; .
\ee
Here and below we use
the $R$-matrix formalism \cite{FRT} with slightly
modified matrix notation \cite{IP1} to simplify the
appropriate calculations. Namely, we use
\be
\lb{3}
\begin{array}{c}
T \equiv T_{1} = T^{i_{1}}_{j_{1}} \delta^{i_{2}}_{j_{2}} \; , \;\;
T' \equiv T_{2} =\delta^{i_{1}}_{j_{1}} T^{i_{2}}_{j_{2}} \; , \\
u \equiv u_{1} = u^{i_{1}}_{j_{1}} \delta^{i_{2}}_{j_{2}} \; ,  \;\;
v' \equiv v_{2} = \delta^{i_{1}}_{j_{1}} v^{i_{2}}_{j_{2}} \; , \\
\R \equiv \hat{R}_{12} = P_{12} R_{12} \; , \;\;
\R' \equiv \hat{R}_{23} = P_{23} R_{23} \; ,
\end{array}
\ee
where $T^{i}_{j}, \; u^{i}_{j}, \; v^{i}_{j} $ are quantum
$N \times N$ matrices;
$\R, \; \R' \in Mat_{N} \times Mat_{N}$ are invertible $R$-matrices;
$P_{12}$ is the permutation matrix and indices
$1,2,3, \dots $ enumerate matrix spaces.
The Yang-Baxter equation for $R$-matrices in this notation reads
$$
R_{12}R_{13}R_{23} = R_{32}R_{13}R_{12} \Leftrightarrow
\R \R' \R = \R' \R \R' .
$$
First of all we recall some known facts
about the algebras (\ref{1}),(\ref{2}) to be used below.
Both algebras (\ref{1}) and (\ref{2})
are the Hopf \cite{FRT} and braided Hopf \cite{Ma3} algebras
respectively. The structure mappings for them are:
\be
\Delta(T^{i}_{j}) = T^{i}_{k} \otimes T^{k}_{j}
\equiv (T \, \tilde{T})^{i}_{j} , \;
{\cal S}(T) = T^{-1}, \; \epsilon(T^{i}_{j}) = \delta^{i}_{j},
\lb{4}
\ee
\be
\Delta(u^{i}_{j}) = u^{i}_{k} \, \underline{\otimes} \, u^{k}_{j}
\equiv (u \, \tilde{u})^{i}_{j}, \;
{\cal S}(u) = u^{-1}, \; \epsilon(u^{i}_{j}) = \delta^{i}_{j} \; ,
\lb{5}
\ee
where $\otimes, \; {\cal S}(T), \; \epsilon(T)$
are the operator tensor product, antipode and counit (see \cite{FRT})
while  $\underline{\otimes}, \; {\cal S}(u), \; \epsilon(u)$
are the braided
tensor product, braided antipode and braided counit
(for their definition see \cite{Ma1}-\cite{Ma3}) .
The braiding for the algebra (\ref{2}),(\ref{5}) is defined by
the relations \cite{Ma1}-\cite{Ma3}:
\be
\lb{6}
\R^{-1} (1 \, \underline{\otimes} \, u) \R (u\, \underline{\otimes}\, 1) =
(u \, \underline{\otimes}\, 1) \R^{-1} (1\, \underline{\otimes}\, u) \R
\Leftrightarrow
\R^{-1} \tilde{u} \R u = u \R^{-1} \tilde{u} \R \; ,
\ee
which specify the braided tensor product $\underline{\otimes}$ in (\ref{5}).
The "braiding" for the quantum groups (\ref{1}),(\ref{4}) is
trivial
$[ T^{i}_{j}, \; \tilde{T}^{k}_{l}] \equiv [ T, \; \tilde{T}'] =0$.

It is well known that the algebra
(\ref{2}) is a covariant comodule
with respect to the adjoint coaction of the quantum group
(\ref{1}),(\ref{4}):
\be
\lb{7}
u^{i}_{j} \rightarrow \Delta_{A}(u^{i}_{j}) =
T^{i}_{k}{\cal S}(T)^{l}_{j} \otimes u^{k}_{l}
\equiv (TuT^{-1})^{i}_{j} \; ,
\ee
where we imply again in the last equality the "trivial braiding":
\be
\lb{8}
 [u, \; T'] = 0 .
\ee
On the other hand, one can find that the algebra (\ref{1})
is a covariant braided comodule with respect to the
left braided coaction of the braided Hopf algebra (\ref{2}),(\ref{5}):
\be
\lb{9}
T^{i}_{j} \rightarrow  \Delta_{B}(T^{i}_{j}) =
u^{i}_{k} \, \underline{\otimes} \, T^{k}_{j} \equiv u \, T ,
\ee
with the nontrivial braiding
\be
\lb{10}
T u' = \R u \R^{-1} T .
\ee
Indeed, $\Delta_{B}(T)$ (\ref{9}) satisfy the R-T-T relations (\ref{1})
in view of eq.(\ref{10}). Then, one can
prove that the comodule axiom
\be
\lb{11}
(\Delta \, \underline{\otimes} \, id)\Delta_{B} =
( id \, \underline{\otimes} \, \Delta_{B})\Delta_{B}
\ee
and the relation (\ref{10}) are consistent
with the braiding (\ref{6}) specific for the
braided Hopf algebra (\ref{2}),(\ref{5}).
Thus, we have demonstrated the interplay of
the quantum groups (\ref{1}) and
the reflection equation (braided) algebras (\ref{2}).
Namely, in addition to the
general phylosophy that the algebras (\ref{1})
and (\ref{2}) are related by the process of transmutation
\cite{Ma2},\cite{Ma3},
we have shown that these two algebras
can be considered as covariant comodules
with respect to the (braided) coactions of one to another.

\vspace{1cm}

{\bf 2.} The main result of this letter is that
an analogous
interplay is inherited for the differential extensions
of the algebras (\ref{1}) and (\ref{2}). Here we consider
the case of linear quantum groups
with the $GL_{q}(N)$ $R$-matrix. In fact,
we need only the Hecke condition for the $R$-matrix:
\be
\lb{11'}
\R^{2} = (q-q^{-1})\R + 1 \; .
\ee

The differential Hopf algebra over $GL_{q}(N)$
(denoted as $\Omega_{A}$)
with the generators $\{ T^{i}_{j}, \; dT^{i}_{j} \}$
is defined by eq.(\ref{1}) and commutation relations
\cite{IP1}-\cite{IsPop}:
\begin{eqnarray}
\hbox{\bf R} (dT) T' & = & T (dT)' \hbox{\bf R}^{-1} \; ,
\lb{12} \\
\hbox{\bf R} (dT) (dT)' & = & - (dT) (dT)' \hbox{\bf R}^{-1} \; .
\lb{13}
\end{eqnarray}
The corresponding structure maps are given by eq.(\ref{4}) and
\cite{Sud},\cite{IsPop},\cite{Brz}
\be
\lb{14}
\begin{array}{c}
\Delta(dT) = dT \otimes T + T \otimes dT \equiv
dT \tilde{T} + T d\tilde{T}, \; \\
{\cal S}(dT) = -T^{-1} dT T^{-1}, \; \epsilon(dT) = 0 \; .
\end{array}
\ee
To define the differential braided Hopf algebra
over $BM_{q}(N)$ (denoted as $\Omega_{B}$)
we have to consider in addition to eq.(\ref{2}) the
following relations \cite{KuA}:
\be
\lb{15}
\begin{array}{c}
\R^{-1} \, u \, \R \, du = du \, \R \, u \, \R , \\
\R \, du \, \R \, du = -du \, \R \, du \, \R^{-1} ,
\end{array}
\ee
and the extension of the comultiplication (\ref{5}) is \cite{IsV}:
\be
\Delta(du) = du \, \underline{\otimes} \, u  +
u \, \underline{\otimes} \, du  \equiv
du \, \tilde{u} + u \, d\tilde{u} .
\lb{16}
\ee
Now our propositions are: \\
{\bf Proposition 1.}
{\it
The differential algebra $\Omega_{B}$
(\ref{2}),(\ref{15})
is a covariant comodule algebra with respect to the following
coaction (homomorphism) of the algebra $\Omega_{A}$
(\ref{1}),(\ref{12}),(\ref{13}):
\begin{eqnarray}
\Delta_{A}(u) & = & T\, u \, T^{-1} \; ,
\lb{17} \\
\Delta_{A}(du) & = & dT\, u \, T^{-1} + T\, du \, T^{-1} +
T\, u \, dT^{-1} \; .
\lb{18}
\end{eqnarray}
The braiding is trivial and defined by eq.(\ref{8}) and

$$
[ du, \; T' ] = [ u, \; dT' ] = [ du, \; dT' ]_{+} =0 \; .
$$
{\bf Proposition 2.}
The differential algebra $\Omega_{A}$
(\ref{1}),(\ref{12}),(\ref{13}) is a covariant braided
comodule with respect to the braided coaction
(homomorphism) of $\Omega_{B}$
(\ref{2}),(\ref{15}):
\begin{eqnarray}
\Delta_{B}(T) & = &  u \, T \; ,
\lb{19} \\
\Delta_{B}(dT) & = & du \, T + u \, dT \; .
\lb{20}
\end{eqnarray}
The braiding is nontrivial and given by eq.(\ref{10}) and
}
\be
\lb{21}
\begin{array}{c}
dT \, u' = \R \, u \, \R^{-1} \, dT \; , \;\;
T \, du' = \R \, du \, \R^{-1} \, T , \\
dT \, du' = - \R \, du \, \R^{-1} \, dT .
\end{array}
\ee

The proofs of Propositions 1 and 2 are straightforward.
For the illustration we verify the homomorphism
(\ref{19}),(\ref{20}) using, for example, the relation (\ref{12})
$$ \R\,\Delta_{B}(dT)\,\Delta_{B}(T')=
\R\,du\,\underline{T \,u'}\,T'+
\R\,u\,\underline{dT \,u'}\,T'= $$
$$=\underline{\R\,du\,R\,u} \,\,\, \underline{\R^{-1}T\,T'}+
\R \,\underline{u\,R\,u\R^{-1}}\,\R^{-1}\,\underline{\R\,dT\,T'}= $$
$$= u\,\underline{\R\,du\,\R^{-1}\,T}T'\,\R^{-1}
 +u\,\underline{\R\,u\,\R^{-1}\,T}\,dT'\,\R^{-1}= $$
$$ =u\,T(du'\,T'+u'\,dT')\,\R^{-1}=
\Delta_{B}(T)\,\Delta_{B}(dT')\,\R^{-1} $$
(underlining indicates the parts to which the next operation is
to be applied). Analogous calculations for eq.(\ref{13}) are in fact
optional because their result can be foreseen from the differentiating the
equality just obtained.

The comodule axiom (\ref{11}) and the braiding relations
(\ref{21}) are consistent with the braiding relations for
the differential algebra over $BM_{q}(N)$ (this braiding and
the comultiplication (\ref{16}) have been proposed by A.A.Vladimirov
and published in \cite{IsV})
\begin{equation}
\label{22}
 \;\;\;\;\;  \left\{
\begin{array}{l}
 \R^{-1}\,\tilde{u}\,\R\,u=u\,\R^{-1}\,\tilde{u}\,\R\,, \\
 \R^{-1}\,d\tilde{u}\,\R\,u=u\,\R^{-1}\,d\tilde{u}\,\R\,, \\
 \R^{-1}\,\tilde{u}\,\R\,du=du\,\R^{-1}\,\tilde{u}\,\R\,,  \\
 \R^{-1}\,d\tilde{u}\,\R\,du=-du\,\R^{-1}\,d\tilde{u}\,\R\, .
\end{array}
\right.
\end{equation}
It can be verified by the substitution $T \rightarrow \tilde{u}T$
into the formulas (\ref{21}).

In the papers \cite{IP2},\cite{IsPop},\cite{Brz},\cite{Aru}
it has been shown that the comultiplication
(\ref{4}),(\ref{14}) for the differential algebra
$\Omega_{A}$ (\ref{1}),(\ref{12}),(\ref{13})
leads to the relation
\be
\lb{23}
\Delta_{A}(\Omega) = \tilde{T} \Omega \tilde{T}^{-1} +
d\tilde{T} \tilde{T}^{-1} \; ,
\ee
where Cartan's 1-forms $\Omega = dTT^{-1}$ satisfy
\be
\lb{24}
\R \, \Omega \R \, \Omega + \Omega \R \, \Omega \, \R^{-1} = 0
\ee
and $[\Omega, \; \tilde{T}'] = 0$. Then, one can introduce the
noncommutative 1-form connections $A$ (transformed as in (\ref{23}))
and the curvature 2-forms $F=dA-A^{2}$
to formulate the so-called quantum group covariant
noncommutative geometry \cite{IsPop}. The same procedure
for the algebra $\Omega_{A}$, but with
the braided coaction (\ref{19}),(\ref{20}),
yields the formula (cf. with (\ref{23}))
\be
\lb{25}
\Delta_{B}(\Omega) = u \, \Omega \, u^{-1} +
du \, u^{-1} .
\ee
Here $u^{-1}$ is a braided antipode introduced
by Sh.Majid in \cite{Ma1}-\cite{Ma3}
and one can deduce from (\ref{21}) the
corresponding braiding relations:
\be
\lb{26}
\begin{array}{c}
\Omega \, \R \, u \, \R^{-1}  = \R \, u \, \R^{-1} \, \Omega , \\
\Omega \, \R \, du \, \R^{-1}  = - \R \, du \, \R^{-1} \, \Omega
\end{array}
\ee
demonstrating the noncommutativity of the "transformation group"
elements $u^{i}_{j}$ and 1-forms $\Omega$.
Now one can again substitute, instead of $\Omega$, the 1-form connections
$A$ transformed as in (\ref{25}) and satisfying relations
(\ref{24}),(\ref{26}). Then the curvature
2-forms $F= dA -A^{2}$ are
transformed homogeneously
\be
\lb{27}
\Delta_{B}(F) = u F u^{-1} .
\ee
The braiding for the operators $F$ and $u$
is deduced from eqs.(\ref{26})
\be
\lb{28}
F \, \R \, u \, \R^{-1}  = \R \, u \, \R^{-1} \, F ,
\ee
\be
\lb{28'}
F \, \R \, du \, \R^{-1}  = \R \, du \, \R^{-1} \, F ,
\ee
and, as it was shown in \cite{Ma2},\cite{Ma3},
the relations (\ref{27}),(\ref{28})
(for arbitrary $R$-matrices) respect the commutation
relations for the curvature 2-forms
\be
\lb{29}
\R \, F \, \R \, F = F \, \R \, F \, \R \; .
\ee
Moreover, relations (\ref{25})-(\ref{28'})
with substitution $\Omega \rightarrow A$ respect the
following cross-commutator for $A$ and $F$:
\be
\lb{30}
\R \, A \, \R \, F = F \, \R \, A \, \R .
\ee
Thus we have the following \\
{\bf Proposition 3.}
{\it The algebra
\begin{eqnarray}
\R \, A \, \R \, A & + &  A \, \R \, A \, \R^{-1} = 0 \; ,
\lb{301} \\
\R \, A \, \R \, F & = & F \, \R \, A \, \R \; ,
\lb{302} \\
\R \, F \, \R \, F & = & F \, \R \, F \, \R \; ,
\lb{303}
\end{eqnarray}
(for the Hecke type $R$-matrix (\ref{11'}))
is a covariant braided comodule algebra with
respect to the braided coaction of $\Omega_{B}$
(\ref{2}),(\ref{15}) (differential extension of $BM_{q}(N)$):
\be
\lb{31}
\Delta_{B}(A^{i}_{j})  =  u^{i}_{k} \, (u^{-1})^{l}_{j}
\, \underline{\otimes} \, A^{k}_{l}  +
du^{i}_{k}\, (u^{-1})^{k}_{j} \, \underline{\otimes} \, 1
\equiv (u\, A\, u^{-1})^{i}_{j} + (du\, u^{-1})^{i}_{j} \; ,
\ee
\be
\lb{32}
\Delta_{B}(F^{i}_{j})  =  u^{i}_{k} \, (u^{-1})^{l}_{j}
\, \underline{\otimes} \, F^{k}_{l}
\equiv (u\, F\, u^{-1})^{i}_{j}  \; .
\ee
The braiding is nontrivial:
\be
\lb{33}
F \, \R \, u \, \R^{-1}  = \R \, u \, \R^{-1} \, F , \;
A \, \R \, u \, \R^{-1}  = \R \, u \, \R^{-1} \, A ,
\ee
\be
\lb{34}
F \, \R \, du \, \R^{-1}  = \R \, du \, \R^{-1} \, F , \;
A \, \R \, du \, \R^{-1}  = - \R \, du \, \R^{-1} \, A ,
\ee
and the comodule axiom (\ref{11}) is consistent with
the braiding relations (\ref{22}). } \\
Propositions 1 and 2 establish the closed relations between
the differential extensions of the quantum groups $\Omega_{A}$
(\ref{1}),(\ref{12}),(\ref{13}) and the reflection equation
(braided) algebras $\Omega_{B}$ (\ref{2}),(\ref{15}).
Proposition 3 shows that the algebra (\ref{301})-(\ref{303})
is covariant not only under the coaction of
$\Omega_{A}$ (see \cite{IsPop}) but also covariant under
the braided coaction of $\Omega_{B}$ (\ref{31}),(\ref{32}).

\vspace{1cm}

{\bf 3.} To conclude this letter we would like to
make some remarks. \\
{\bf A.)} As it has been pointed out above,
the algebra (\ref{301})-(\ref{303})
considered in Proposition 3 has the geometrical interpretation
when the generators $A, \; F$ are associated via the
relation $F=dA-A^{2}$.
Namely, in this case, one can consider $A$ as 1-form connections
while $F$ as curvature 2-forms.
Now we introduce the braided adjoint co-invariants
using the well known q-trace \cite{FRT},\cite{Res}
\be
\lb{35}
C_{2k} = Tr_{q}(F^{k}) = Tr(DF^{k}),
\ee
where the matrix $D$ is related to the $R$-matrix
(see e.g. \cite{Res},\cite{KS},\cite{Ma2},\cite{Zum}):
$$
{D^{i}}_{j}= { \tilde{R}^{ki} }_{jk} =
Tr_{(2)} \left( P_{12} ((R_{12}^{t_{1}})^{-1})^{t_{1}} \right)
$$
and $\tilde{R} = ((R_{12}^{t_{1}})^{-1})^{t_{1}}$.
By definition, the $2k$-forms $C_{2k}$ (\ref{35}) are
co-invariants not only under the adjoint coaction of
the quantum groups (\ref{1}) (see \cite{IsPop}) but also
under the braided co-transformations (\ref{32}). Moreover,
these $2k$-forms commute with $A$ and $F$ and, as
it has been shown in \cite{IsPop}, they are closed:
\be
\lb{36}
dC_{2k} = Tr_{q}(AF^{k} - F^{k}A) = 0 \; .
\ee
To prove the last equality in (\ref{36}) one has to use
the commutation relations (\ref{302}).
Therefore, the central elements $C_{2k}$
could be interpreted as noncommutative analogs
of the Chern characters. \\
{\bf B.)} One can generalize Proposition 2 in the following way.
Let us consider the differential algebra
$\overline{\Omega}_{B}$ generated by
$\{ v^{i}_{j}, \; dv^{i}_{j} \}$ (cf. with (\ref{2}),(\ref{15})):
\be
\lb{37}
\begin{array}{c}
\R^{-1} \, v' \, \R^{-1} \, v' =
v' \, \R^{-1} \, v' \, \R^{-1} , \\
 \R^{-1} \, v' \, \R^{-1} \, dv' = dv' \, \R^{-1} \, v' \, \R , \\
 \R^{-1} \, dv' \, \R^{-1} \, dv' = - dv' \, \R^{-1} \, dv' \, \R .
\end{array}
\ee
This algebra is a differential braided Hopf algebra with
the comultiplication
\be
\lb{37'}
\begin{array}{c}
\Delta(v) = v \, \underline{\otimes} \, v \equiv v \, \tilde{v}  \; , \\
\Delta(dv) = dv \, \underline{\otimes} \, v  +
v \, \underline{\otimes} \, dv  \equiv
dv \, \tilde{v} + v \, d\tilde{v} ,
\end{array}
\ee
and braided relations (cf. with (\ref{22}))
\begin{equation}
\label{37''}
 \;\;\;\;\;  \left\{
\begin{array}{l}
 \R \,\tilde{v}'\,\R^{-1}\,v'=v'\,\R\,\tilde{v}'\,\R^{-1}\,, \\
 \R\,d\tilde{v}'\,\R^{-1}\,v'=v'\,\R\,d\tilde{v}'\,\R^{-1}\,, \\
 \R\,\tilde{v}'\,\R^{-1}\,dv'=dv'\,\R\,\tilde{v}'\,\R^{-1}\,,  \\
 \R\,d\tilde{v}'\,\R^{-1}\,dv'=-dv'\,\R\,d\tilde{v}'\,\R^{-1}\, .
\end{array}
\right.
\end{equation}
Then, we have the following: \\
{\bf Proposition 4.}
{\it The differential algebra $\Omega_{A}$
(\ref{1}),(\ref{12}),(\ref{13}) is a covariant braided
comodule with respect to the braided coaction
(homomorphism) of two commuting algebras $\Omega_{B}$
(\ref{2}),(\ref{15}) and $\overline{\Omega}_{B}$
(\ref{37})
$
( [ \Omega_{B} , \; \overline{\Omega}_{B} ]_{\pm} = 0 ).
$
This coaction of $\Omega_{B} \otimes \overline{\Omega}_{B}$
can be represented in the form:
\begin{eqnarray}
\Delta_{LR}(T) & = &  u \, T \, v \; ,
\lb{38} \\
\Delta_{LR}(dT) & = & du \, T \, v + u \, dT \, v +
u \, T \, dv \; .
\lb{39}
\end{eqnarray}
The braiding is defined by eqs.(\ref{10}),(\ref{21})
and }
\be
\lb{40}
\begin{array}{c}
 v \, T' =  T' \, \R \, v' \, \R^{-1} \; , \;\;
 v \, dT' = dT' \, \R \, v' \, \R^{-1}  \; , \\
dv \, T' = T' \, \R \, dv' \, \R^{-1} \; , \;\;
dv \, dT' = - dT' \, \R \, dv' \R^{-1} \; .
\end{array}
\ee
The proof of this Proposition is the same as the proof of
Proposition 2.

Propositions 1-4 lead us to the natural conjecture that
the differential algebras $\Omega_{A}$ (\ref{1}),(\ref{12}),
(\ref{13}) and $\Omega_{B}$ (\ref{2}),(\ref{15}) are
related by the process of transmutation considered by
Sh.Majid in \cite{Ma4},\cite{Ma2},\cite{Ma3}. \\
{\bf C.} There are some arguments that the braided
quantum group covariant noncommutative geometry
(briefly discussed in Proposition 3 and in the subsection A.))
could be associated with the global version
of the $q$-gauge theories proposed by
L.Castellani \cite{Cas}. For example,
the matrix elements $u^{i}_{j}$ generating "gauge transformations"
(\ref{31}),(\ref{32}) do not commute with the "$q$-gauge fields"
$A$ and $F$ (see \cite{Cas} and (\ref{33}),(\ref{34})).
It would be very interesting to trace these relations
completely by means of formulating the "infinitesimal
version" of eqs.(\ref{31}),(\ref{32}).

\section*{Acknowledgments} The author would like to thank
L.Castellani, P.N.Pyatov and A.A.Vladimirov for
helpful discussions.

This work was supported in part by the
Russian Foundation of Fundamental Research (grant 93-02-3827).

\end{document}